# Giant magnetostriction in Tb-doped $Fe_{83}Ga_{17}$ melt-spun ribbons

**Wei Wu, Jinghua Liu, Chengbao Jiang\* and Huibin Xu**

School of Materials Science and Engineering, Beihang University, Beijing 100191, P.R. China

Giant magnetostriction is achieved in the slightly Tb-doped $Fe_{83}Ga_{17}$ melt-spun ribbons. The tested average perpendicular magnetostriction $\lambda_\perp$ is -886 ppm along the melt-spun ribbon direction in the $Fe_{82.89}Ga_{16.88}Tb_{0.23}$ alloy. The calculated parallel magnetostriction $\lambda_\parallel$ is 1772 ppm, more than 4 times as large as that of binary $Fe_{83}Ga_{17}$ alloy. The enhanced magnetostriction should be attributed to a small amount of Tb solution into the A2 matrix phase during rapid solidification. The localized strong magnetocrystalline anisotropy of Tb element is suggested to cause the giant magnetostriction.

Giant magnetostrictive materials have the potential applications in actuators, sensors and energy harvesting devices. [1-3] The substitution of nonmagnetic Ga into common body centered cubic Fe enhanced the magnetostriction $\lambda$ over tenfold, which has drawn lots of attentions. However, the reported magentostriction of Fe-Ga alloys is 400 ppm, just one fifth of that of TbDyFe alloys. [4-9] Focusing on improving the magnetostriction of Fe-Ga alloys, many efforts have been made by doping the third elements. [10-17] Doping the interstitial elements B, C and N slightly increases the magnetostriction of Fe-Ga alloys. [11, 12, 17] Alloying with the $3d$ and $4d$ transition elements, including Ni, V, Cr, Mn, Co, Mo, Rh, etc drastically decreases the magnetostrictive strain of Fe-Ga alloys. [10, 13-15] Doping the one main group elements Si, Ge and Sn also cannot enhence the magnetostriction. [16] Up to now, it is not obvious to improve the magnetostriction of Fe-Ga alloys by doping the above third elements.

The magnetostriction of ferromagnetic materials originates from the spin-orbit coupling, with the same origin of the magnetocrystalline anisotropy. [18, 19] The relatively low magnetostriction of Fe-Ga alloys stems from the low magnetocrystalline anisotropy. The giant magnetostriction 2600 ppm for $TbFe_2$ alloy originates from its strong magnetocrystalline anisotropy, which is two orders higher than that of Fe-Ga alloys. [4-9] We believe that the localized strong magnetocrystalline anisotropy possibly induces a giant magnetostriction. Based on this idea, the third element Tb is selected to dope into the $Fe_{83}Ga_{17}$ alloy to form the localized strong magnetocrystalline anisotropy. However, there is no solid solution of Tb in the A2 phase of $Fe_{83}Ga_{17}$ alloy. Melt spinning is expected to increase the solid solution of Tb into the A2 matrix of $Fe_{83}Ga_{17}$ alloy by the rapid solidification. In this letter, slight amount of Tb element is doped in $Fe_{83}Ga_{17}$ alloys by melt spinning. The tested average perpendicular magnetostriction $\lambda_\perp$ is -886 ppm along the melt-spun ribbon direction in the $Fe_{82.89}Ga_{16.88}Tb_{0.23}$ alloy. The calculated parallel magnetostriction $\lambda_\parallel$ is 1772 ppm, more than 4 times as large as that of binary $Fe_{83}Ga_{17}$ alloy.

High-purity starting elements iron, gallium and, terbium with the purity of 99.99% were arc melted under argon atmosphere for four times with the nominal compositions of $Fe_{83}Ga_{17}$, $Fe_{82.97}Ga_{16.97}Tb_{0.06}$, $Fe_{82.89}Ga_{16.88}Tb_{0.23}$ and $Fe_{82.77}Ga_{16.76}Tb_{0.47}$, named as Tb0, Tb0.06, Tb0.23 and Tb0.47 respectively. Ribbons were prepared by melt spinning with the copper wheel velocity of 20 m/s. The crystal structure was characterized by using x-ray diffraction (XRD) on a Rigaku X-ray diffractometer with Cu $K\alpha$ radiation. The composition and morphology of the ribbons were determined using a JEOL JXA-8100 electron probe micro-analyzer (EPMA) and a JEOL JEM-2100 transmission electron microscope (TEM). Initial magnetization curves (*M-H*) were measured on physical properties measurement system (PPMS). The Curie temperature was detected by a NETZSCH TSA449 thermogravimetry analyzer (TG) with the 20 ℃/min heating rate from room temperature to 1000 ℃. The magnetostriction was tested along the ribbon direction by standard resistance strain gauge technique. The ribbon was pended vertically and the magnetic field was applied perpendicular to the ribbon plane. The upper part of the ribbon was fixed with clamp and the bottom part was loaded with a mass block in order to avoid the bending of ribbons during the magnetization processes. The magnetostriction was measured three times for each sample with the strain gauge stuck on both sides of the ribbons. And then, rotating the ribbons by 180°, the magnetostriction was measured for another three times.

Figure 1 shows the back scattered electron (BSE) and bright-field images of the ribbon samples. All the ribbons exhibit columnar crystal morphology with the average grain size of 10 μm in Fig. 1 (a). Single phase is observed for the Tb0, Tb0.06 and Tb0.23 ribbons, and dual-phase microstructure exists for Tb0.47 ribbon as shown in Fig. 1(b). The second globular phase precipitates in the Tb0.47 ribbon. EDS tests confirm that these second phases are Tb-rich phases, with the composition of $Fe_{60}Ga_{30}Tb_{10}$, indicating the solubility limit of Tb atoms in $Fe_{83}Ga_{17}$ alloy lower than 0.47 at% under the rapid solidification.

X-ray diffractions patterns for the Tb-doped $Fe_{83}Ga_{17}$ ribbons are shown in Fig. 2. All the detected diffraction peaks are indexed as (110), (200), (211) and (220) of the body-centered cubic (bcc) structure, same as A2 phase in the pure $Fe_{83}Ga_{17}$ alloy. The small quantity of the precipitated phase cannot be detected by x-ray diffraction.

The lattice parameters $a$ of the ribbons are calculated according to the XRD data, as plotted in Fig. 3(a). With increasing Tb content, the lattice parameters first increase from $a = 0.2900$ nm for Tb0 to $a = 0.2904$ nm for Tb0.23, and then trend to saturation with $a = 0.2904$ nm for Tb0.47. The initial increase of the lattice parameter $a$ should be associated with the doping of the big Tb atoms into A2 matrix, and the saturation of the lattice parameter $a$ should be attributed to the precipitation of the second Tb-rich phase. This result is consistent with the x-ray diffraction patterns measurement, as shown in Fig.2.

Figure 3 (b) shows that the saturation magnetizations $M_S$ of the ribbon samples first increase as the Tb content increase from 180.0 emu/g for Tb0 to 183.6 emu/g for

Tb0.23 and then trend to saturation with 183.9 emu/g for Tb0.47. The result shows the ferromagnetic coupling of Tb element doped into A2 matrix, different from the usual ferrimagnetic coupling between Fe and Tb. As can be seen from Fig. 3 (c), the Curie temperatures $T_C$ of the ribbon increases first and then slightly decrease with the increase of Tb content. The initial increase of the Curie temperature further confirms the ferromagnetic coupling interactions of the Tb doping into the A2 matrix. The precipitation of the Tb-rich phase results in the decrease of the Curie temperature. The enlargement of the lattice parameter, the enhancement of saturation magnetization, and the increase of the Curie temperature of the Tb-doped Fe-Ga alloys reflect the Tb element solution in the A2 matrix.

The room temperature perpendicular magnetostrictions ($\lambda_\perp$) are measured along the length direction of the ribbon samples and also normal to the direction of the applied field. Each ribbon sample has been tested for twelve times, as mentioned above, the tested data, accompanied with the average value of the magnetostrictions, are plotted in Fig. 4 (a). One of the $\lambda$-$H$ curves for each ribbon is also presented in Fig. 4 (b). For the binary $Fe_{83}Ga_{17}$ (Tb0) ribbon, the average magnetostriction value $\lambda_\perp$ is -176 ppm, consistent with the previously reported value of -163ppm for $Fe_{81}Ga_{19}$ ribbon [20, 21]. The average magnetostriction values of the ribbon sample obviously increase from -176 ppm for Tb0, -447 ppm for Tb0.06 to -886 ppm for Tb0.23 and slightly decease to -862 ppm for Tb0.47. The results indicate that the giant magnetostriction is achieved by slight amount of Tb element doping in the $Fe_{83}Ga_{17}$ ribbon.

The parallel magnetostrictions $\lambda_{//}$ along the direction of the applied field, vertical to the plane of the ribbons, cannot be measured directly since the ribbons are too thin. Therefore, the parallel magnetostrictions $\lambda_{//}$ have been calculated. The relationship of the measuring magnetostriction and the saturation magnetostriction $\lambda_s$ follows the Equation (1), which is based on the isotropic grains. [22]

$$\bar{\lambda} = \frac{3}{2}\lambda_s\left(\cos^2\theta - \frac{1}{3}\right) \quad (1)$$

There, $\theta$ is the included angle between the measuring direction and magnetizing direction as shown in Fig. 4 (a). The columnar grains, normal to the plane of the ribbons, are anisotropic, but they are isotropic in the plane of the ribbons. The tested perpendicular magnetostriction $\lambda_\perp$ along the length direction of the ribbon stems from the in-plane isotropy. It is reasonable to calculate the parallel magnetostrictions $\lambda_{//}$ based on the following Equation (1). Here the saturate magnetostriction $\lambda_s$ equals to the parallel magnetostriction $\lambda_{//}$ and the tested average magnetostriction $\bar{\lambda}$ equals to the perpendicular magnetostriction $\lambda_\perp$. The angle $\theta$ between the measuring direction and the magnetizing direction is 90° as shown in Fig. 4(a). If we substitute $\theta$=90° into Eq. (1)

$$\lambda_\| = -2\lambda_\perp \quad (2)$$

A giant magnetostriction of 1772 ppm is achieved in the Tb0.23, which is more than 4 times larger than the reported value of 400 ppm in the Fe-Ga alloys, and comparable to that of the giant rare earth alloys Terfenol-D, where the amount of rare earth (Tb, Dy) is more than 60 %.[2, 23]

The magnetostriction and the magnetocrystalline anisotropy of Fe-Ga alloys commonly originate from the spin-orbit coupling. Due to their relatively low magnetocrystalline anisotropy, the Fe-Ga alloys exhibit a moderate magnetostriction of about 400 ppm. The rare earth element Tb with high magnetocrystalline anisotropy is doped into Fe-Ga ribbons, forming the localized strong magnetocrystalline anisotropy. A giant magnetostriction of 1772 ppm is achieved in the Tb0.23 ribbon.

Therefore, it is believable that the localized strong magnetocrystalline anisotropy can also induce a giant magnetostriction.

This work is supported by National Basic Research Program (2012CB619404)，Natural Science Foundation of China（50925101, 51221163, 91016006）.

Fig.1 (a) BSE images of the nominal compositions of Tb0, Tb0.06, Tb0.23 and Tb0.47 ribbon samples. (b) Bright-field images of the Tb0.23 ribbon with single phase microstructure and Tb0.47 ribbon with the globular precipitate phases.

Fig.2 XRD patterns of melt-spun ribbons for Tb0, Tb0.06, Tb0.23 and Tb0.47.

Fig.3 Composition dependence of the lattice constant, saturation magnetization $M_s$ and Curie temperature $T_C$ for the Tb-doped $Fe_{83}Ga_{17}$ ribbons.

Fig.4 (a) Magnetostriction of Fe Tb0, Tb0.06, Tb0.23 and Tb0.47 ribbons. The inset shows the measuring direction and magnetizing direction. (b) One of the $\lambda$-$H$ curves for each ribbon samples.

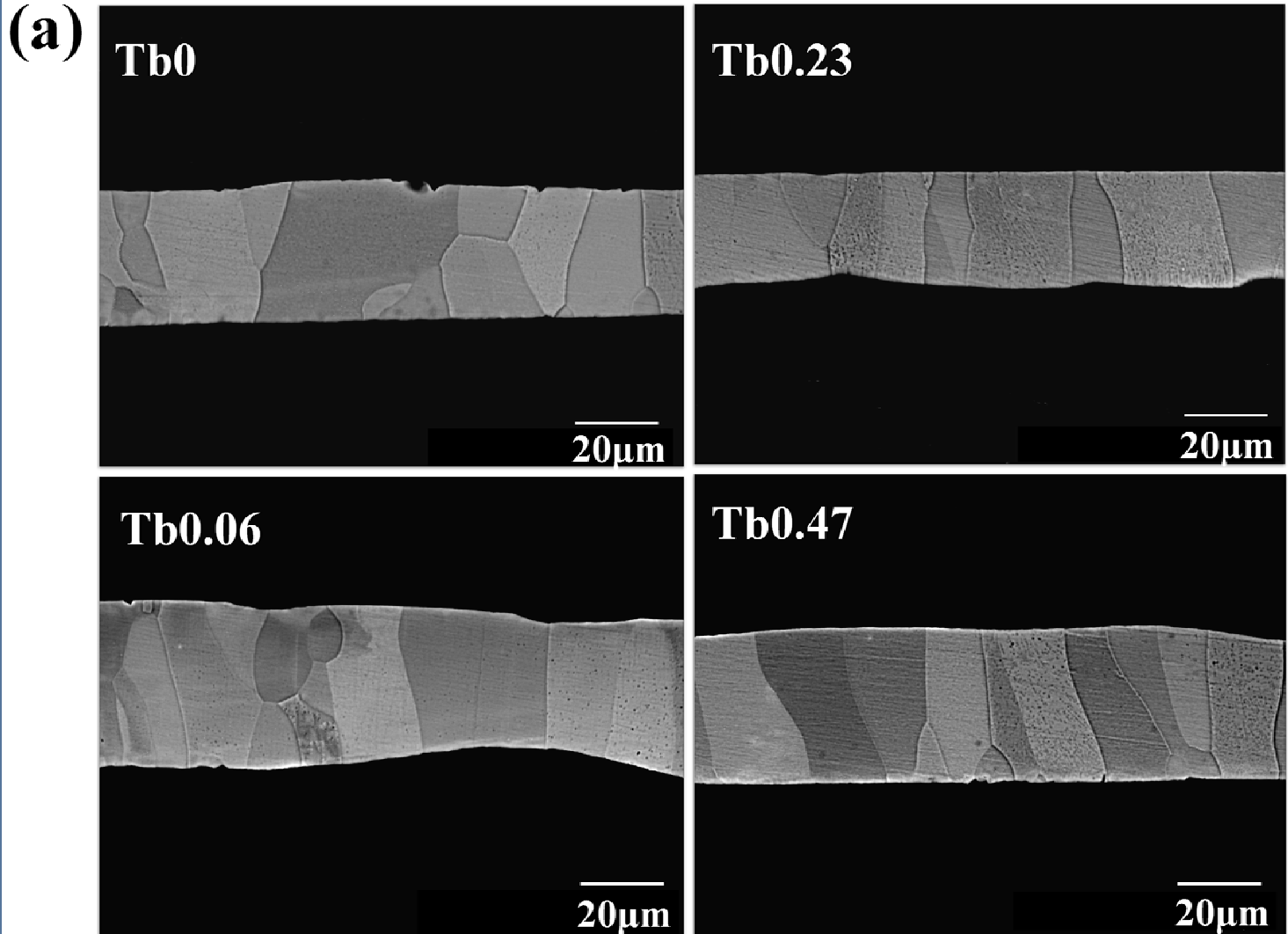
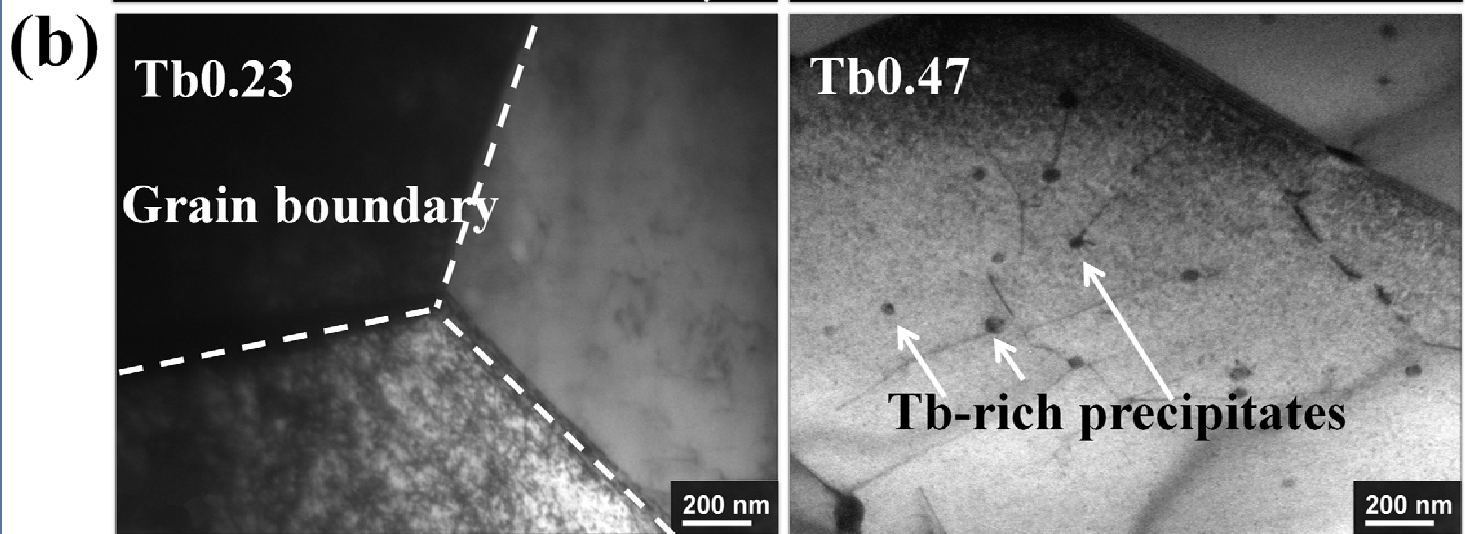

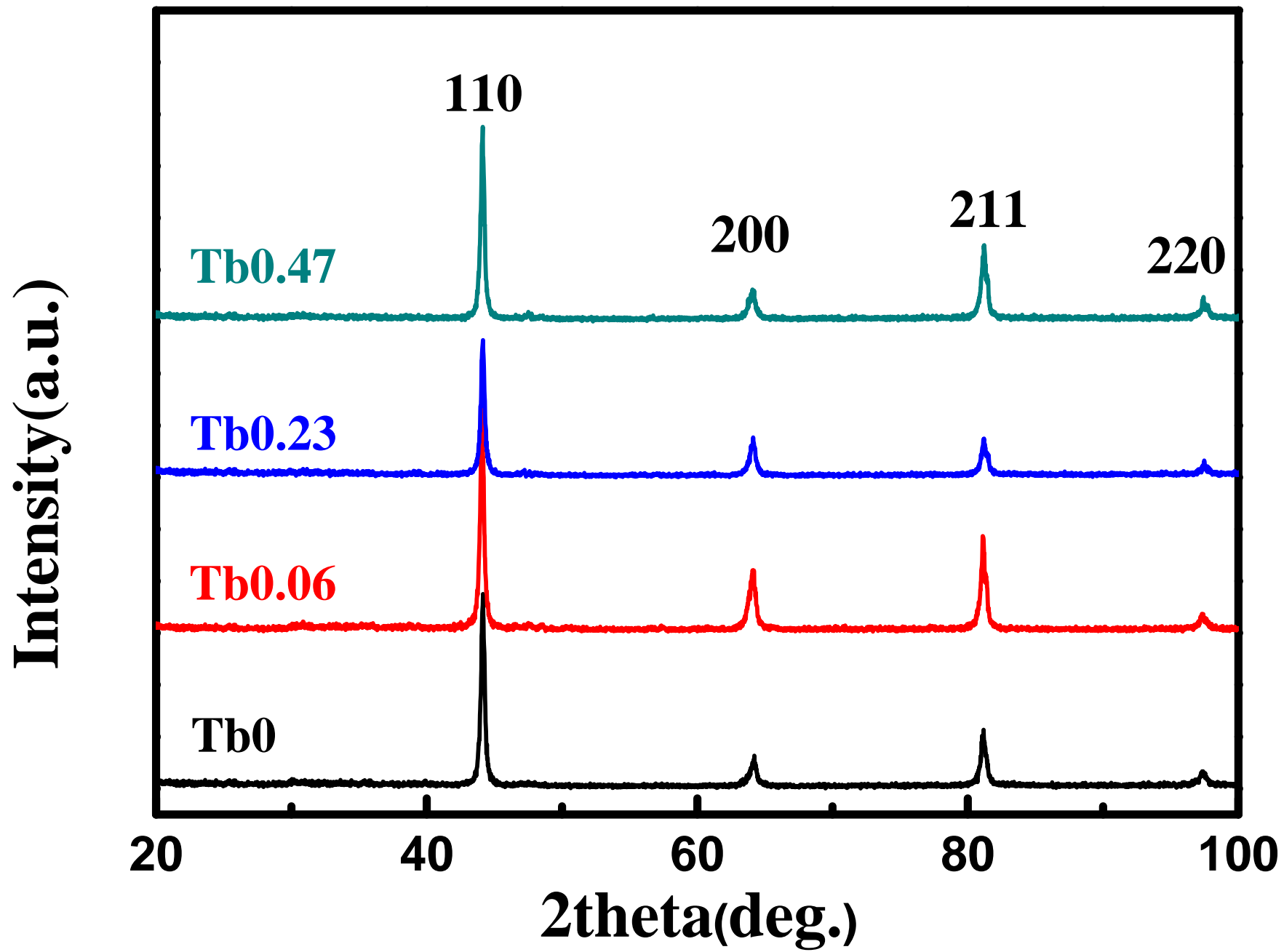

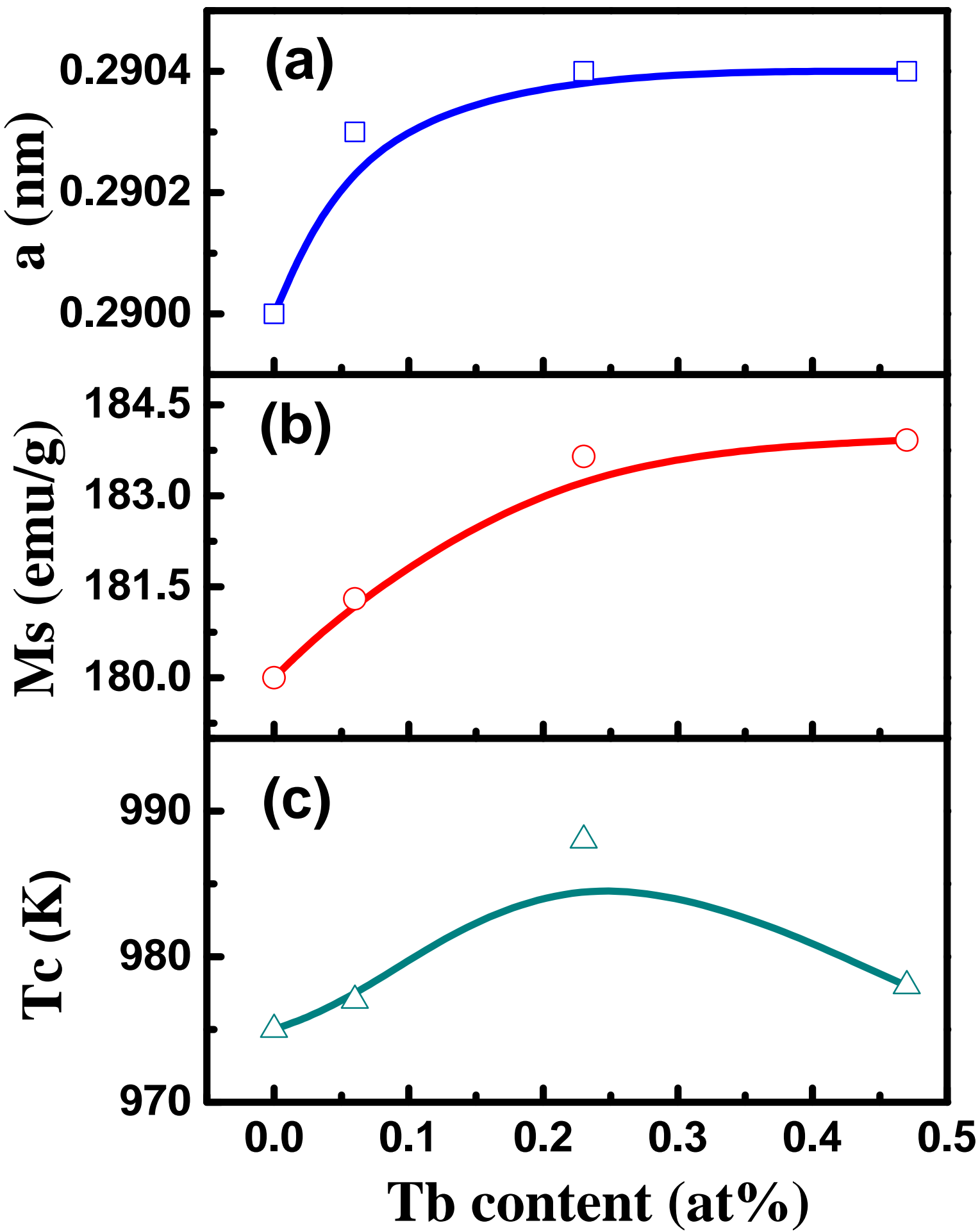

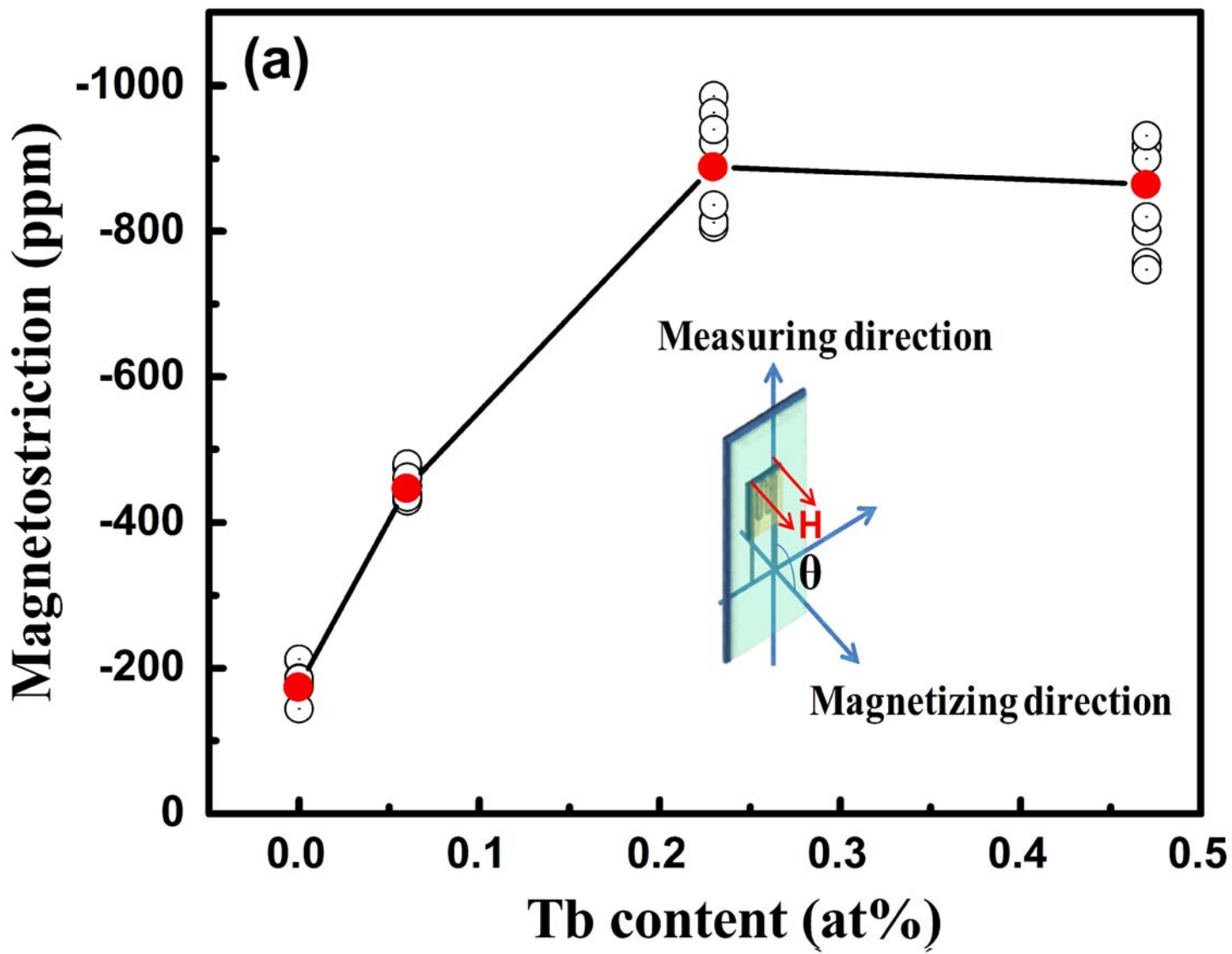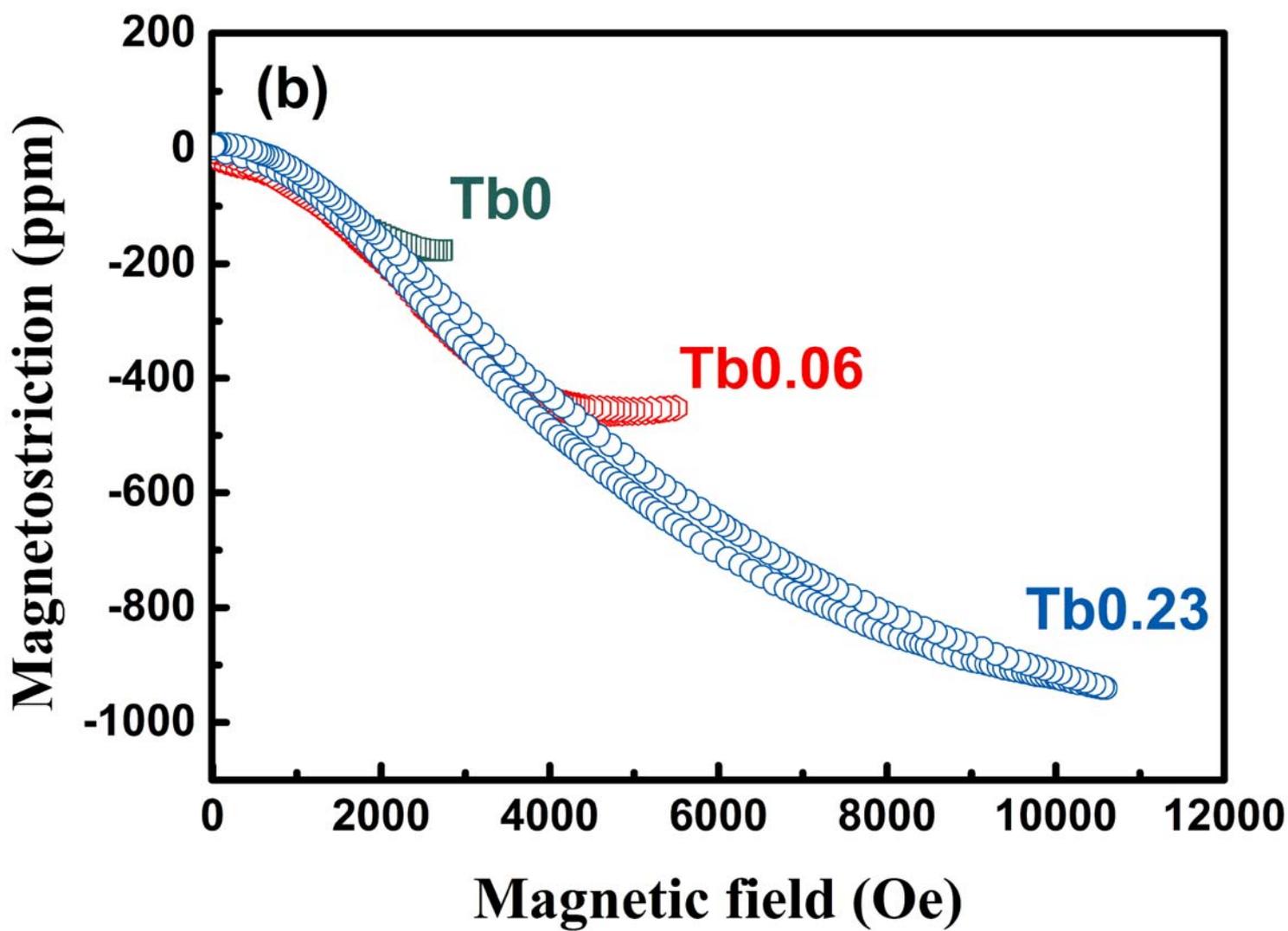